# The challenges and opportunities of delivering wireless high speed broadband services in Rural and Remote Australia: A Case Study of Western Downs Region (WDR)

**Sanjib Tiwari**
Australian Digital Futures Institute
University of Southern Queensland
Queensland, Australia
Email: sanjib.tiwari@usq.edu.au

**Michael Lane**
School of Management and Enterprises
University of Southern Queensland
Queensland, Australia
Email: michael.lane@usq.edu.au

**Khorshed Alam**
School of Commerce
University of Southern Queensland
Queensland, Australia
Email: khorshed.alam@usq.edu.au

## Abstract

This paper critically assesses wireless broadband internet infrastructure, in the rural and remote communities of WDR in terms of supply, demand and utilisation. Only 8 of 20 towns have ADSL/ADSL2+, and only 3 towns have 4G mobile network coverage. Conversely all of the towns have 2G/3G mobile network coverage but have problems with speed, reliability of service and capacity to handle data traffic loads at peak times. Satellite broadband internet for remote areas is also patchy at best. Satisfaction with existing wireless broadband internet services is highly variable across rural and remote communities in WDR. Finally we provide suggestions to improve broadband internet access for rural and remote communities. Public and private investment and sharing of wired and wireless broadband internet infrastructure is needed to provide the backhaul networks and 4G mobile and fixed wireless services to ensure high speed, reliable and affordable broadband services for rural and remote communities.

## 1  INTRODUCTION

Still broadband internet penetration and use in rural and remote area of developed countries such as USA, UK, Canada and Australia has been much lower compared to the urban areas of these countries (ABS 2014; Dwivedi et al. 2010b; Grubesic 2012; Kawade and Nekovee 2012; Prieger 2013; Rajabiun and Middleton 2013). An assessment of existing telecommunication network infrastructure in rural and remote communities reveals a low level of broadband penetration with many black spots where telecommunication infrastructure do not exist. A combination of poor telecommunications infrastructure coverage, low population density, inadequate regulation, and a focus by telecommunications companies on high-cost technologies designed for urban markets makes Internet connectivity in many parts of the rural and remote area a complex and costly proposition (Arai et al. 2012; Grubesic 2012; Hill et al. 2014; Kawade and Nekovee 2012). Due to inadequate information and communication infrastructure and facilities in rural and remote area, the digital divide between rural and remote communities and urban communities is widening (Park et al. 2015; Willis and Tranter 2006).

To overcome this situation, in USA American Federal Communication Commission has driven the uptake of faster rates of data transfer practically anywhere, either over 4G cellular wireless or over satellite (Titch 2013). Similarly, European countries such as Germany, France, Ireland, UK have utilized mix of broadband technology (wired and wireless LTE technology such as 4G/ WiMAX) to optimize the cost of the broadband networks and maximise the reach to the rural and remote households to make broadband a feasible offering in sparsely populated areas (EC 2014).

The purpose of this study is to evaluate the wireless broadband infrastructure in rural and remote Australia in terms of supply, demand and utilisation. To the authors knowledge this study is the first attempt to develop an in-depth profile of wireless broadband supply demand and utilisation in rural and remote Australia. Researchers believe that this case study of Western Downs Region (WDR) could be extended to other rural and remote parts of the Australia to obtain more complete picture of the wireless broadband infrastructure and digital divide that exists within rural and remote areas due to variability and availability of broadband technologies providing broadband connectivity. The research was conducted in two phases, using a mixed methods approach in an in-depth case study of WDR. Firstly a descriptive evaluation of the digital infrastructure available in study area was conducted, followed by a large scale of survey of households in rural and remote communities in WDR. This study provides a clearer picture of current wireless broadband internet infrastructure in rural and regional Australia particularly WDR in terms of access, speed and availability. This study also captured households' attitudes and perceptions current wireless broadband internet services in comparison to wired broadband internet services. Finally this study suggests emerging wireless technology solutions which might provide suitable wireless broadband connectivity technologies for rural and remote communities.

The structure of this paper is as follows. First the paper begins with review of the current state of broadband in Australia. Then the paper looks at the wireless broadband technologies which are deployed in Australia including Mobile broadband internet and satellite broadband. The data collection method is subsequently explained. Findings are analysed and policy implications are discussed for improving last-mile broadband connection in rural and remote Australia using wireless broadband technologies. Finally concludes with a summary and outlines possible implication of the study.

## 2 LITERATURE REVIEW

### 2.1 Broadband Internet definition

Broadband technology includes a variety of high-speed access technologies that include Fibre Optic, ADSL/ADSL2+, cable modems, satellite and wireless networks (3G, 4G). The term broadband does not have an established definition, it varies in different nations and is defined on basis of transmission and technology so that yesterday's fast speed broadband is today's narrow band (Choudrie and Dwivedi 2006). However, for the purpose of this study we follow OECD and National Broadband Network (NBN) and they define broadband as high-speed, always on, internet connectivity, with a minimum download speed of 256 kbps or greater and minimum upload speed of 64 kbps (NBN 2014a; OECD 2015). In reality current speed of wireless and fixed wired broadband internet is much faster than these and is measured in mbps rather than kbps. In this paper we are focusing on wireless broadband technologies because these are the broadband access technologies most widely available in rural and remote Australia.

### 2.2 Wireless Broadband Internet

Wireless broadband is high-speed Internet and data service delivered through a wireless local area network or wide area network. As with other wireless service, wireless broadband may be either mobile or fixed. A fixed wireless service such as WiMAX provides wireless Internet connectivity to user specific or permanent locations, such as homes and offices (Middleton and Bryne 2011). Mobile broadband services are delivered via wireless networks and enhance connectivity of users of advance technologies (such as smart phones and other wireless devices) any place at any time (Gulati et al. 2015; Srinuan et al. 2012).

#### 2.2.1 Mobile Broadband Internet

In Australia there are three major mobile broadband network providers: Telstra, Optus and Vodafone. Telstra one of the most available mobile broadband in terms of coverage of area provide Next G (3G/4G) services. Optus second largest mobile network is used by several other providers (such as TPG, Amaysim, Boost, Dodo, Virgin Mobile), while Vodafone is used by Crazy John's (and Red Bull, but it doesn't offer a data-specific service) (Kidman 2012). The existing mobile network operators have spectrum holdings from 850MHz to 2 GHz these include 1800 MHz (held by Telstra, Optus and Vodafone), 850 MHz (held by Telstra and Vodafone), 900 MHz (held by Telstra, Optus and Vodafone) and 2 GHz (held by Telstra, Optus and Vodafone). These are in most cases fully utilised providing 2G and 3G mobile services.

Current Australian government has the following framework for mobile and fixed wireless broadband network deployment at rural and remote level:
- The NBN provides a national wholesale access product for telecoms retail service providers. This includes a fixed wireless TD-LTE network under construction that will cover approximately 600,000 premises (NBN 2015a). NBN Co's current specifications use the 2.3GHz spectrum band. Its Cell Coverage Area is defined as being where a single end-user can achieve at least 25Mbps downlink and 5Mbps uplink with 95 percent probability (NBN 2014b).
- NBN Co has further access to used digital dividend spectrum 694-820MHz and unused spectrum 3.4GHz band (ACMA 2013; NBN 2014b)

### 2.2.2 Satellite Broadband Internet

NBN Co launched the Interim Satellite Service (ISS) on 1 July 2011 to provide an improved service to rural and remote Australians ahead of the rollout of the Long Term Satellite Service. NBN Co's Interim Satellite service (using rented Optus and IPStar) reached capacity in December 2013 after 48,000 premises ordered a service (NBN 2015c). To be able to provide fast broadband to rural and remote areas of Australia, NBN Co is launching two new dedicated Long Term Satellites to reach 3% of rural and remote customers and will be able to offer wholesale services configured for a planned 25 megabits per second (Mbps) download and 5Mbps upload (Locke 2015; NBN 2015b).

## 2.3 Demand for Wireless broadband Access Spectrum in Australia

The demand for mobile and fixed wireless access service spectrum (WAS) in Australia is continuing to grow. Figure 1 shows the take-up of internet services by technology type, based on the Internet Activity Survey conducted by the ABS in December 2012. This growth in mobile broadband services is problematic for the mobile network operators – because the demand for data is growing exponentially but the returns on the investment do not justify the investment into improved mobile broadband infrastructure in rural and remote Australia because the economies of scales cannot be delivered because of low population densities. In many cases mobile network operators need to lease backhaul capacity to meet the exponential growth in data transmission with little in the way of substantially increased revenue streams. So this presents a situation where government intervention in the broadband services markets may be necessary in order to guarantee reliable and high speed mobile broadband access in rural and remote Australia.

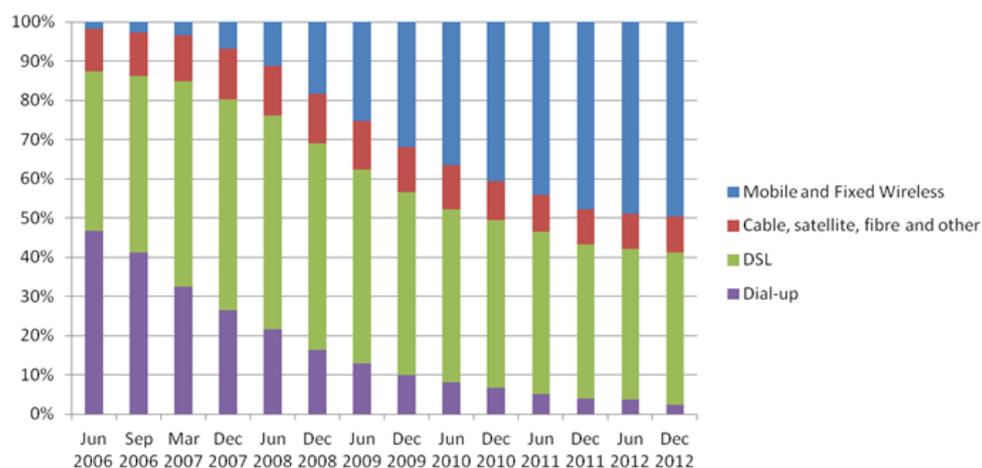

*Figure 1: Uses of spectrum by different internet services by technology type, source ACMA (2013)*

As a result of increased data traffic on mobile networks, demand for additional WAS in the market to has been increasing, but this is tempered somewhat by the evolution and improvement of mobile 3G and 4G technologies including HSPA and LTE. For example, additional efficiencies are now being realised in the 1800 MHz band with the transition from GSM to LTE and aggregation of spectrum into larger contiguous bandwidths. Telstra will decommission its GSM mobile network on December 1 2016 and redeploy the 2G spectrum for 3G and 4G mobile network services (Coyne 2014). This demand growth is already placing pressure on existing mobile networks and will necessitate continued network deployments in the foreseeable future.

## 2.4 Wireless Broadband Internet in rural and remote Australia

Current mobile wireless technologies have limited capacity and coverage in rural and remote communities and satellite technologies are expensive and provide limited data rates. A total of 890 MHz of spectrum is allocated, or planned to be allocated, to WAS in Australia, though not available in all regions and the spectrum used is limited to below 4 GHz. So even though the adoption and use of broadband internet has increased substantially over time in Australia, when looking at the macro data it does not provide a true picture of the digital divide that actually exists between urban Australia and rural and remote Australia. Rice and Katz (2003) defined digital divide is the gap between those who use digital technologies such as broadband Internet and those who do not understood how to use these digital technologies or do have access to these digital technologies and in binary terms distinguishing the "haves" from the "have-nots".

## 2.5 Theoretical Framework - Mobile Broadband Ecosystem

The World Bank Ecosystem Framework for deployment and adoption of Broadband provided the theoretical and conceptual framework for understanding the supply (penetration), demand (adoption) and utilisation (absorptive capacity and productive use based on prior knowledge, experience and skills) of Wireless Broadband services in WDR (Cohen and Levintbal 1990; Kelly and Rossotto 2012; World Bank 2012). The productive use of wireless broadband is restricted by the lack of mobile broadband infrastructure capacity and investment in rural and remote Australia. The supply (penetration) of wireless broadband services is fundamental to ensuring rural and remote communities in regions such as WDR are sufficiently covered with fast and reliable broadband services to enable industries and communities to adopt and utilise the benefits of the digitalisation of the economy both at national and international level.

## 3 METHODOLOGICAL APPROACH

The following research questions were systematically addressed in a case study of WDR: Q1. What is comparative availability of wired and wireless broadband internet services for rural and remote communities in WDR? Q2. What are the perceptions of households of rural and remote communities in WDR comparatively of wired and wireless broadband internet services?

This research conducted an explanatory case study for the evaluation of the supply, demand and use of broadband internet in rural and remote communities in Western Downs Region. An explanatory case study allowed us to focus on gaining an in-depth of understanding of the phenomena of the study the impact of digital infrastructure supply on household adoption and use of broadband services in rural and remote communities (Yin 1994, Harder 2004). A mixed methods approach was used to collect to multiple sources (Gable 1994; Venkatesh et al. 2013). A mixed methods approach using multiple sources of data allowed the researchers to capture a richer picture and deeper understanding of the phenomena of study. Rural and remote communities are classified in this study based on Rural, Remote and Metropolitan Areas (RRMA) classification where urban centres fall within the ranges of 10,000 – 24,999 population, and 5000 to 9,999 (rural urban centres), 100 to 4999 (remote urban centres) (AIHW 2004). Majority of the small towns in rural and remote Australia are using either cellular wireless network (3G or 4G) or satellite internet. To determine and assess broadband infrastructure available in WDR archival data from the ABS and RFNSA (www.rfnsa.gov.au) (RFNSA 2014) and publicly contributed social media websites such as www.adsl2exchange.com.au (ADSL2 Exchanges 2013) were used. The study also conducted tests on cellular broadband speed using the OpenSignal mobile app for the three Mobile network services available in Western Downs Region (WDR).

This research used a large scale survey of households in WDR to collect quantitative data about household adoption and use of broadband internet. In this paper we report on household adoption and their perceptions of use of wireless and wired broadband comparatively. The survey instrument was developed from a number of previous studies of broadband technology adoption which have used instruments proven to be valid and reliable (Straub et al. 2004). The survey instrument was pretested with a number of academics with extensive experience in survey research design. The survey was then piloted with a number of households before the main survey data collection was undertaken. The survey instrument is available on request from the authors. The printed version of the survey was distributed in person to randomly selected households using a stratified sampling method to ensure that the 1500 surveys were distributed to a sample representative of the population localities in WDR. The targeted respondent was the major decision maker in each household. Three hundred and one usable survey responses were obtained. A response rate of 20 percent was achieved which is comparable to similar previous studies of broadband adoption and use by households (Brown and Venkatesh 2005; Dwivedi et al. 2010a; Dwivedi et al. 2006).

## 4   BACKGROUND AND DEMOGRAPHICS OF WDR

The data were collected in Western Downs Region, Queensland Australia, and is considered to be representative of rural and remote Australia. In this study we consider rural areas as a major populated locations such as (Dalby, Chinchilla, Tara, Miles, Jandowae and Wandoan) and define remote areas as rest of the WDR (Warra, Jimbour, Bell, Condamine, Meandarra, Brigalow, Dulacca etc.). The region WDR has a covers a land area of 38,039 square kilometres and contains a population of 32,000 making it one of the top 20 largest councils in Queensland in terms of area. WDR has six major shires: Dalby, Chinchilla, Miles, Jandowae, Tara and Wandoan. Approximately 60% of the population resides in the main towns, with some of these being quite small. Most of the town-based population is located in Dalby (58%) and Dalby and Chinchilla combined account for almost 80% of the town-based population in WDR. In this study 206 (68.4%) respondents are from rural area and 95 (31.6%) respondents from remote part of WDR. The respondent have an average family annual income is $60,000 which is below the state average household annual income $68,000. The education levels of respondents were as follows 45.5% had a secondary school education, 22.6% have higher education, 16.3% have a have diploma, 9% have just primary level of education. One of the interesting finding by Park (2012) in US that broadband users have higher level of income and higher level education. In this study, 80% of the respondent own laptop and notebooks followed by 64% own some kind of smart phone, 58% own desktop and 55% own tablets to use Internet. Furthermore, 72.4% of the ADSL/ADSL2+ and 70% of wireless broadband users were homeowners and 27.5% ADSL/ADSL2+ and 30% of wireless broadband users were live in renting or share with others.

### 4.1   Results of Analysis of Mobile Broadband Ecosystem in WDR

#### 4.1.1   Existing Broadband Infrastructure in WDR

From the World Bank (2012) theoretical framework we can see that broadband connection or demand is strongly influenced by the existing network infrastructure or supply (see Table 1 for digital infrastructure coverage in WDR).

|              | Pop.  | PSTN | ADSL | ADSL2+ | Fibre optic backbone | 3G | 4G | Satellite broadband | Proprietary Radio networks |
|---|---|---|---|---|---|---|---|---|---|
| Dalby        | 16000 | Y | Y | Y | Y | Y | Y | ALC | ALC |
| Chinchilla   | 7200  | Y | Y | Y | Y | Y | Y | ALC | ALC |
| Miles        | 1450  | Y | Y | Y | Y | Y | Y | ALC | ALC |
| Jandowae     | 1000  | Y | Y | Y | N | Y | N | ALC | ALC |
| Wandoan      | 400   | Y | Y | Y | N | Y | N | ALC | ALC |
| Bell         | 300   | Y | Y | N | N | Y | N | ALC | ALC |
| Condamine    | 135   | Y | Y | Y | N | Y | N | ALC | ALC |
| Tara         | 1100  | Y | Y | Y | N | Y | N | ALC | ALC |
| Meandarra    | 250   | Y | Y | Y | N | Y | N | ALC | ALC |
| Kaimkillenbun| 100   | Y | N | N | N | Y | N | ALC | ALC |
| Macalister   | 15    | Y | N | N | N | Y | N | ALC | ALC |
| Brigalow     | 40    | Y | N | N | N | Y | N | ALC | ALC |
| Kogan        | 40    | Y | N | N | N | Y | N | ALC | ALC |
| Dulacca      | 120   | Y | N | N | N | Y | N | ALC | ALC |
| Drillham     | 70    | Y | N | N | N | Y | N | ALC | ALC |
| Moonie       | 40    | Y | N | N | N | Y | N | ALC | ALC |
| Glenmorgan   | 20    | Y | N | N | N | Y | N | ALC | ALC |
| Warra        | 120   | Y | N | N | N | Y | N | ALC | ALC |
| Jimbour      | 40    | Y | N | N | N | Y | N | ALC | ALC |
| Gulugaba     | 10    | Y | N | N | N | N | N | ALC | ALC |

*Table 1. Mapping of Digital Infrastructure coverage in Western Downs Region*
Legend Y = Available; N = Not Available; ALC = Available in Limited Circumstances

The supply of broadband infrastructure in Australian rural and remote communities like those in WDR has some interesting characteristics. Firstly Telstra is the major telecommunication provider for the whole of WDR. Telstra has a monopoly on the PSTN which underpins the delivery of ADSL/ADSL2+ services. There are other ISPs providing ADSL/ADSL2+ services in WDR but in many of the smaller towns if ADSL/ADSL2+ is available Telstra are the only provider of ADSL/ADSL2+ services. Telstra also has the most comprehensive mobile network coverage (2G/3G/4G) although 4G Mobile services are currently only available in the largest towns. Furthermore Optus and Vodafone have progressively limited mobile network coverage which is restricted to the largest towns in WDR. However the seemingly comprehensive 2G/3G mobile network coverage as indicated in Table 1 is misleading because mobile network coverage is patchy at best once you leave the perimeter of a town. The predominant 3G mobile network coverage in smaller towns is often the only broadband internet service available which can easily degrade because it is a shared bandwidth with limited capacity. Outsides these towns' rural and remote communities are forced to use either mobile networks (2G/3G) or satellite connection. There are a few fibre optic networks in WDR such as NextGen Group and Nexium Networks which is a commercial subsidiary of Ergon. These provide an extensive fibre optic backhaul network for key population areas populations (Dalby, Chinchilla and Miles) in WDR but this fibre network is only accessible by network services wholesalers and local government. There has been a similar occurrence with mining companies such QGC – BG group which rolled out their own private fibre optics network to connect all liquefied natural gas plants and control their operation remotely from head office in Brisbane. These mining companies have extensive mobile data networks but these are private commercial networks with limited access given only to government for emergency services..

### 4.1.2 Broadband Adoption and Use by Households in WDR

The results of the survey indicate that three quarters of households have been using the Internet for six or more years. Interestingly almost 40% of households started using the Internet more than 16 years ago when the primary means of accessing the Internet was via a dialup modem. The majority of households (76%) are accessing the internet on a daily basis. A further 13%of households are accessing the internet on a weekly basis while only 1% of households are accessing the internet on a less than weekly basis. This would indicate that the adoption and demand for Broadband services is significant across the respondent households in WDR. In this study, 97% of survey respondents said most common digital service used by their households for communication (Email, Instant messaging, VoIP) access followed by 94 % information seeking (information search, news, maps), 78% for e-commerce activities (pay bills, banking, auctions, shopping), 71% for Social media and networking (Facebook, twitter, LinkedIn, Instagram, blog, forum), 68% entertainments activities (music, games, videos, gambling) and 62% e-organizational services (education services, health services, employment services, government services. Table 2 shows how rural and remote households in WDR are accessing the Internet.

| Type of Internet connection | Rural (n=206) | Remote (n=95) |
| --- | --- | --- |
| ADSL/ADSL2+ (Wired Broadband) | 125 (86.2%) | 20 (13.8%) |
| 3G/4G Wireless broadband | 116 (64.1%) | 65 (35.9%) |
| Satellite wireless broadband | 6 (20.7%) | 23 (79.3%) |
| Dial-up | 3 (60.0%) | 2 (40.0%) |

*Table 2. Type of Internet connection by rural and remote households in WDR*

Results summarised in Table 2 shows that 125 (86%) of rural communities households were connected through fixed line wired connection such as ADSL/ADSL2+ whereas only 20 (14%) of remote households were connected by fixed wired connection. Similarly the number of wireless broadband 3G/4G connections for rural households 116 (64%) were very high compare to remote households where only 65 (36%) have 3G/4G wireless broadband connections. We could see there is a significant difference in percentage of households with wireless broadband 3G/4G connections between rural and remote households. This could be because of their households' location and availability and reliability of wireless broadband 3G/4G services in those locations. We could see that almost 23 (80%) of remote household are using satellite wireless broadband. According to many of the respondents comments that satellite broadband reception is patchy at best. Remote communities have little or no access to wired broadband hence they are reliant on wireless mobile broadband or satellite broadband internet services which are highly variables in terms of access, speed and quality of services.

According Puschita et al. (2014) user satisfaction is influenced by the quality of the broadband services provided and the level of expectation. In our survey in order to capture rural and remote households attitudes to broadband in WDR we asked them to rate a number of questions related to continued use, satisfaction with current broadband internet service, cost of broadband internet, and intention to upgrade to NBN or higher speed broadband internet when available. A seven point likert scale was used where 1 meant "Strongly Disagree" and 7 meant "Strongly Agree". The results are shown in Table 3, with means and standard deviation scores for each of these questions.

| WDR Household Attitudes to Broadband | Mean | Standard deviation | Attitude |
|---|---|---|---|
| **Continue using broadband internet in the future** | 6.32 | 0.966 | Positive |
| **Upgrading to NBN or higher speed broadband internet when its available** | 6.00 | 1.183 | Positive |
| **Satisfied with our current broadband internet service and plan** | 4.39 | 1.942 | Relatively Positive |
| **Subscribing to broadband internet is quite expensive** | 4.36 | 1.768 | Relatively Negative |

*Table 3. Rural and remote users' perception of Broadband Internet*

Table 3 results shows that rural and remote households' attitudes towards continued use of broadband Internet services are generally favourable. A score of 6.32 for continued use of broadband internet indicate that rural households are strongly disposed to continued use of broadband Internet and they would like to upgrade their broadband internet connection to a NBN higher-speed broadband connection when it's available. The rural and remote households are relatively positive and satisfied with the Internet services and plans they are currently using. However there is significant variability in their satisfaction with Broadband Internet services with a SD (± 1.942) that shows there are a number of unsatisfied users as well. Some of the satisfied respondent indicated in the survey that "they are lucky to have a broadband connection in their region and their thoughts were with those far remote people who do not have any type of internet connection". This may be the reason why many households seem happy with what they do have. However, this study also suggests that rural and remote households are not happy with cost of Broadband internet services. It might be because of the price difference between urban and rural area for broadband internet access is substantial. Some of the survey respondents make the comments that "why we are paying more money for less speed and less data as compared to city people". Most of the broadband Internet services are expensive in rural area in terms of data download/upload when compared to urban areas (see whistleout.com.au or broadbandguide.com.au/).

The above findings indicate that rural and remote households positively predisposed toward broadband internet services and the adoption and demand for broadband services is significant across the households in WDR despite variable availability and quality of service of wired and wireless broadband internet across the region. To generate a more accurate profile of the network technology most households are using to access Internet in rural and remote areas, we conducted a cross tabulation analysis of the survey data in terms of their satisfaction with the various broadband internet access technologies summarised in Table 4.

|  |  | Percentage of households Dissatisfaction | Strongly Disagree | Disagree | Somewhat Disagree | Neither Agree nor Disagree | Somewhat Agree | Agree | Strongly Agree | Percentage of households Satisfaction |
|---|---|---|---|---|---|---|---|---|---|---|
| **3G/4G** | Rural | 28% | 12% | 5% | 11% | 22% | 11% | 23% | 15% | 49% |
|  | Remote | 50% | 21% | 12% | 17% | 9% | 14% | 20% | 6% | 40% |
|  | Overall | 36% | 15% | 8% | 13% | 17% | 12% | 22% | 12% | 46% |
| **Satellite** | Rural | 17% |  |  | 17% | 17% | 17% | 50% |  | 67% |
|  | Remote | 56% | 26% | 13% | 17% | 9% | 17% | 9% | 9% | 35% |
|  | Overall | 48% | 21% | 10% | 17% | 10% | 17% | 17% | 7% | 41% |
| **ADSL/ ADSL2+** | Rural | 18% | 3% | 6% | 9% | 17% | 19% | 28% | 18% | 65% |
|  | Remote | 35% | 10% | 5% | 20% | 15% | 20% | 20% | 10% | 50% |
|  | Overall | 20% | 4% | 6% | 10% | 17% | 19% | 27% | 17% | 63% s |
| **Dial-up** | Rural | 50% |  |  |  |  |  |  | 100% | 100% |
|  | Remote |  |  | 50% | 50% |  |  |  |  | 0% |
|  | Overall | 40% |  | 20% | 20% |  |  |  | 60% | 60% |

*Table 4. Level of satisfaction of Internet services by rural and remote households*

The level of satisfaction with fixed wired broadband such as ADSL broadband services is much higher (60%) compared to wireless 3G/4G Mobile broadband services (46 percent) and Satellite broadband services (41%). This is not surprising given that ADSL/ADSL2+ Broadband services are faster, have bigger data quotas and are significantly cheaper comparatively to Mobile Broadband services. Moreover if you look more deeply at this data, rural households are more satisfied 49% and 65% respectively with 3G/4G and ADSL/ADSL2+ as compared to remote households with 40% satisfaction with 3G/4G services and 50% satisfaction with ADSL/ADSL2+. This finding suggest that 3G/4G Mobile broadband services and Satellite broadband services are much more variable in terms of coverage, speed and reliability of service than fixed line ADSL broadband services in remote areas compared to rural areas hence the lower levels of satisfaction in remote households. Furthermore, the speed of 3G/4G mobile broadband services is really patchy in both rural and remote areas and 4G is only available in the 3 largest towns in WDR. ADSL2+ is only available in 8 of the 20 towns in WDR. Dial-up which is classified as a narrowband internet service has very low usage these days and our findings confirm this similar to recent reports by the ABS on Internet access technologies.

We conducted speed tests using open signal app in three different mobile (Telstra, Optus and Vodafone) network provider in WDR, over one month's period in Feb-March 2014, across the three major large towns and a number of small towns

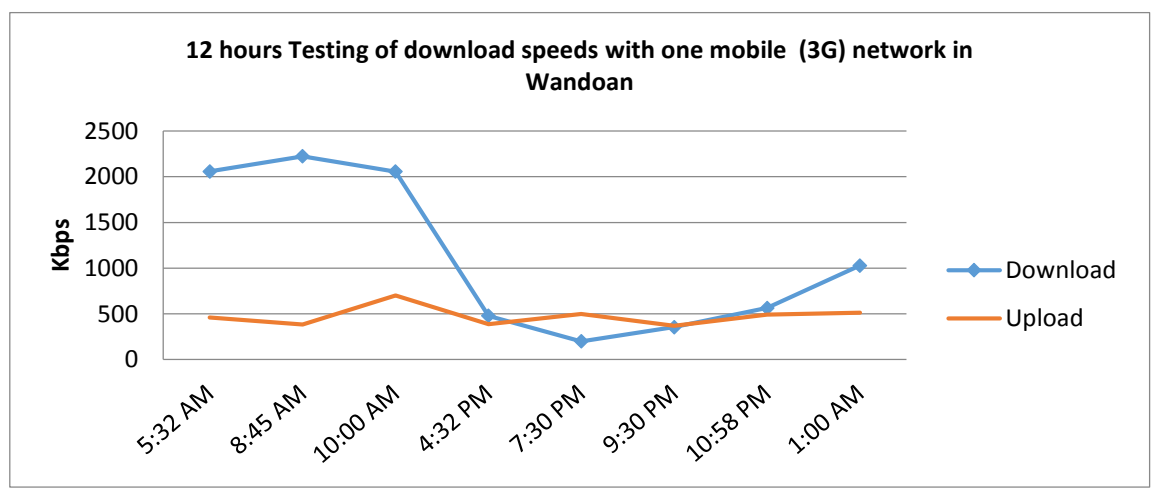

*Figure 3: Testing of download speeds with Optus mobile (3G) network in Wandoan*

In one remote town Wandoan, the OpenSignal speed test results show (Figure 3) that broadband download/upload speed for one carrier degrade in peak usages times however this testing is not conclusive. Results indicate that when many people are using wireless mobile broadband they are not able to utilize the broadband internet services because the network speed degrades. Hence there is a need to improve the wireless network technology to provide these rural and remote people with quality of service in terms of improved internet speeds and reliability of services.

# 5 IMPROVING WIRELESS BROADBAND INTERNET EFFECTIVENESS IN RURAL AND REMOTE LOCALITIES IN WDR

There could be great opportunities for private and public sectors to join together to build a robust and affordable broadband infrastructure for rural and remote communities. Hence, we need to embrace the full opportunities of fixed wireless broadband and mobile broadband technology convergence in rural and remote area Optic fibre technology though expensive has been proven to deliver the best bandwidth for broadband service delivery. The purposed wireless access solutions are adapted based on WDR internet users profile, geographical nature of rural and remote areas and a broadband market analysis in different areas by Puschita et al. (2014), Talukder et al. (2013) and Islam et al. (2008).

| **Technology** | **3G+UMTS/HSDPA/ Mobile carrier 4G** | **Fixed wireless 4G(LTE/WiMAX)** |
|---|---|---|
| **Operation** | License required (ACMA Australia) | License required (ACMA Australia) |
| **Infrastructure** | **Complex (terminals, base station subsystem, network and switching subsystem)** | **Simple (subscriber stations and base stations)** |
| **Speed** | **Medium (elaborate cell planning and frequency allocation)** | **Fast (minimal cell planning and frequency allocation)** |
| **Area Coverage** | Wide area (small and large cells ranges – hundreds of meters to tens of km) | Wide area (small and large cells ranges – hundreds of meters to tens of km |
| **Cell Capacity** | Very high (hundreds to thousands of users per cell) | Very high but fixed number (hundreds to thousands of users per cell) |
| **Data rates** | Medium (hundreds of Mbps) | high data rates (hundreds of Gbps) |
| **User terminals** | dedicated devices (provider-linked subscriber module) | Dedicated devices (operator looked subscriber module) |
| **Deployment costs** | High costs (hundreds of thousands of dollars per base station) | High costs (hundreds of thousands of dollars per base station) |
| **Terminal costs** | High (tens to hundreds of dollars per terminal) | High (tens to hundreds of dollars per terminal) |
| **Subscription cost** | High (dedicated subscription required, traffic limitations and associated traffic costs, tens to hundreds of dollars per month) | High (dedicated subscription required, no traffic Limitations, tens to hundreds of dollars per month) |

*Table 5. Comparison the main parameters for consideration in network deployment adopted Source (Puschita et al. 2014; Talukder et al. 2013)*

Table 5 shows some of the initials requirements and comparison among the latest wireless networking technology which are widely used for last mile connections around the world. Technologies such as 3G+UMTS/HSDPA/ Mobile carrier 4G have benefits which include contention-free channel access method, a controlled level of any interface over the area, an implicitly higher cell capacity and greater coverages, however these networking technologies need to share the certain bandwidth spectrum among users and number of users is not restricted in a cell. Whereas on the other hand fixed wireless 4G (LTE/WiMAX) have limited number of users assigned for each cell so that they can guarantee the speed, reliability and capacity. A fixed wireless service differs from mobile wireless networks 'as a fixed service is designed to support a dedicated number of premises, each with a wireless receiver affixed externally to the property. This enables the delivery of a more predictable level of service performance to premises compared to a mobile wireless network which is focussed on mobility'. Talukder et al. (2013) compared the cell radius obtained in the case of LTE and WiMAX under different environments and frequencies. The comparison results revealed that the LTE technology not only covers greater area than WiMAX for similar frequencies but also provides high-speed, better Quality of Services and "all-IP" mode of communications. As the results, it is highly expected to be the best candidate for future wireless communication system in rural and remote areas.

As we can see from Table 5 and Puschita et al. (2014) and Talukder et al. (2013) studies the initial cost associated with both technologies is similar. It is not attractive for a mobile network operator to upgrade their cell towers to fixed wireless 4G(LTE/WiMax) technology at the same time because they had already invested and deployed 3G+UMTS/HSDPA technology at considerable cost. The cost of building mobile cell towers is significant and can be between $A100,000 and $A200,000 and requires road and power access. However it would be the fastest and most reliable wireless broadband technology for rural and remote communities.

Nationwide fibre optic backbone development would lead to sustainable development for the provision of high speed last mile connectivity in rural and remote area. However, complete deployment of fibre itself in the rural and remote areas may be commercially less viable because population densities in rural and remote communities in Australia does not justify ROI (return of investment) cost of FTTH/FTTN rollout. For financial viability, and to reduce the wholesale price for fixed wireline broadband capacity NBN Co decided to make fixed wireless broadband available in rural and remote Australia. Here the policymakers need to take a more holistic approach to solving this problem, and part of that is using the NBN to overcome the very high costs of regional backhaul. There is a need to explore the option to review the integration of the fixed wireless and the private sector regional mobile networks that often use common towers and backhaul. Moreover, there are number of dark fibre cable networks laid by private companies for their business. There would be great opportunities to improve broadband internet access and capacity for rural and remote households if these dark fibre optic networks could be used as a backhaul service to support greater connectivity of fixed wireless and 4G networks.

## 6　CONCLUSIONS AND IMPLICATIONS

The provision of adequate infrastructure for rural and regional Australia is clearly an important factor in maintaining competitiveness of Australian rural and remote economy. All rural and remote communities need universal access to broadband to ensure economic prosperity, social inclusiveness, and general networking between communities. The challenge of providing high-speed access to rural and remote areas is one that needs to be tackled both now.

Today's broadband is far better for urban communities compared to the 1.6 million Australians living in rural and remote regions who often have poor and unreliable internet connection. If government did not take further action regarding high-speed broadband then rural and remote communities will be left further behind in the digitalisation of the economy. The rural and remote areas of the WDR suffer from variable and limited telecommunication infrastructure coverage which means that mobile phones often have no reception outside towns. Mobile coverage is poor and internet services are only available in the main settlement areas. There is no opportunity to access or severely limited access to Broadband internet in a number of locations. This area is seen as a priority given the community's increasing expectations of being able to access the latest wireless telecommunication technology for the last mile connectivity.

Our study reveals that the number of households using wireless broadband (60%) overall is much higher than the number of households using fixed wired broadband in WDR (50%). This also suggest that wireless technologies are well established to meet the increased demand from rural and remote households to access internet services for different purposes. Results also suggest that a number of households using wireless broadband are satisfied with the wireless broadband services available to them. However there are significant numbers of households particular in remote areas which are not satisfied with their current wireless broadband internet connection. Moreover they would like to update their wireless broadband services to high speed wireless broadband services when it is available in their locality. In rural and remote areas this could be better if the Australian Federal government would provide the investment needed for the fibre optic networks required for backbone connectivity throughout rural and remote centres. The digital dividend spectrum 694-820 MHz can provide fixed wireless 4G broadband using LTE services in addition to improved long term broadband satellite services for hard to reach locations. It could be more economically feasible to roll out high speed wireless broadband if better use is made of a mix of technology solutions and existing private wireless and dark fibre optic networks where it is available. This would offer the most affordable wireless broadband internet services and reasonable data rates with an inherent advantage to quick deployment in these rural and remote regions of Australia.

## 7　REFERENCES

ABS. 2014. "8153.0 - Internet Activity, Australia, December 2014." Canberra: ABS.

ACMA. 2013. "New Frequency Ranges for Wireless Microphones." http://acma.gov.au/Citizen/Consumer-info/All-about-spectrum/Wireless-microphones/new-arrangements-for-wireless-microphones Retrieved 25th October, 2015.

ADSL2 Exchanges. 2013. "Adsl2 Exchanges." http://www.adsl2exchanges.com.au/ Retrieved 25th June, 2015.

AIHW. 2004. "Rural, Regional and Remote Health: : A Guide to Remoteness Classifications." http://www.aihw.gov.au/WorkArea/DownloadAsset.aspx?id=6442459567 Retrieved 15 November, 2015.

Arai, Y., Naganuma, S., and Satake, Y. 2012. "Local Government Broadband Policies for Areas with Limited Internet Access. An Analysis Based on Survey Data from Japan," *Netcom. Réseaux, communication et territoires*:26-3/4), pp. 251-274.

Brown, S., and Venkatesh, V. 2005. " Model of Adoption of Technology in Households: A Baseline Model Test and Extension Incorporating Household Life Cycle," *MIS Quarterly* (29:3), pp. 399-426.

Choudrie, J., and Dwivedi, Y. K. 2006. "Investigating Factors Influencing Adoption of Broadband in the Household," *Journal of Computer Information Systems* (46  4), pp. 25-34.

Cohen, W. M., and Levintbal, D. A. 1990. "Absorptive Capacity: A New Perspective on Learning and Innovation," *Administrate Science Quarterly* (35:1), pp. 128-152.

Coyne, A. 2014. "Telstra Prepares to Shut Down 2g Network." http://www.itnews.com.au/News/390220,telstra-announces-closure-of-2g-network.aspx Retrieved July, 2015.

Dwivedi, Y., Alsudairi, M., and Irani, Z. 2010a. " Explaining Factors Influencing the Consumer Adoption of Broadband," *International Journal of Business Information Systems* (5:4), pp. 393-417.

Dwivedi, Y. K., Alsudairi, M. A., and Irani, Z. 2010b. "Explaining Factors Influencing the Consumer Adoption of Broadband," *International Journal of Business Information Systems* (5:4), pp. 393-417.

Dwivedi, Y. K., Choudrie, J., and Brinkman, W. P. 2006. "Consumer Usage of Broadband in British Households," *International Journal of Services and Standards* (2:4), pp. 400-416.

EC. 2014. "Engage- Development of High Speed Broadband Networks and Services Throughout Rural Europe." http://ec.europa.eu/digital-agenda/en/news/engage-development-high-speed-broadband-networks-and-services-throughout-rural-europe Retrieved October, 2015.

Francis, H. 2015. "Australian 4g Download Speeds Slump," in: *The Sydney Morning Herald* Sdney: The Sydney Morning Herald

Gable, G. 1994. " Integrating Case Study and Survey Research Methods: An Example in Information Systems," *European Journal of Information Systems* (3:2), pp. 112-126.

Grubesic, T. H. 2012. "The Us National Broadband Map: Data Limitations and Implications," *Telecommunications Policy* (36:2), pp. 113-126.

Gulati, G., Huang, J., Marabelli, M., Weiss, J., and Yates, D. 2015. "Determinants of Mobile Broadband Affordability: A Cross-National Comparison," in: *21st Americas Conference on Information Systems,*. Puerto Rico.

Hill, S. R., Troshani, I., and Burgan, B. 2014. "Broadband Adoption in Regional and Urban Households," *Journal of Computer Information Systems* (54:3), pp. 57-66.

Islam, A. R., Selvadurai, N., and Town, G. 2008. "Wireless Broadband Technologies for Regional and Rural Australia: A Last-Mile Perspective," *Telecommunications Journal of Australia* (58:2-3), pp. 28.21-28.18.

ITU. 2010. "The Digital Dividend - Opportunities and Challenges." https://www.itu.int/net/itunews/issues/2010/01/27.aspx Retrieved October, 2015.

Kawade, S., and Nekovee, M. 2012. "Is Wireless Broadband Provision to Rural Communities in Tv Whitespaces Viable? A Uk Case Study and Analysis," *IEEE International Symposium on Dynamic Spectrum Access Networks (DYSPAN),* , Bellevue, WA, USA: IEEE, pp. 461-466.

Kelly, T., and Rossotto, C. M. 2012. *Broadband Strategies Handbook*. Washington, D.C.: World Bank Publications.

Kidman, A. 2011. "Ultimate Hotspot Shows Telstra Next G Is Feeling the Strain." http://www.lifehacker.com.au/2011/07/ultimate-hotspot-shows-that-telstra-next-g-is-feeling-the-strain/ Retrieved October, 2015.

Kidman, A. 2012. "Planhacker: Every Australian Prepaid Mobile Broadband Deal Compared." http://www.lifehacker.com.au/2012/07/planhacker-every-australian-prepaid-mobile-broadband-deal-compared/ Retrieved 27 September, 2015.

Locke, S. 2015. "Nbn Co's Satellites Promise Internet Speeds a Hundred Times Faster Than Current Dial-up for Regional Australia." http://www.abc.net.au/news/2015-02-09/nbn-groundstations-ready-for-satellite-launch/6075626 Retrieved July, 2015.

McKell, W. 2013. "Superfast Broadband : The Future Is in Your Hands." NSW, Australia: The McKell Institute.

Middleton, C. A., and Bryne, A. 2011. "An Exploration of User-Generated Wireless Broadband Infrastructures in Digital Cities," *Telematics and Informatics* (28:3), pp. 163-175.

NBN. 2014a. "Broadband." http://www.nbnco.com.au/glossary.html Retrieved June, 2015.

NBN. 2015a. "Fixed Wireless Broadband: A Global Comparison." http://www.nbnco.com.au/content/dam/nbnco2/documents/ovum-fixed-wireless%20-broadband-a-global-comparison.PDF Retrieved July, 2015.

NBN. 2015b. "Long Term Satellite Service." http://www.nbnco.com.au/connect-home-or-business/information-for-home-or-business/satellite.html Retrieved July, 2015.

NBN. 2015c. "The Nbn™ Satellite Support Scheme." *Broadband Service Locato* https://programs.communications.gov.au/NBNBSL/ Retrieved July, 2015.

OECD. 2015. "Broadband Access." https://data.oecd.org/broadband/wireless-mobile-broadband-subscriptions.htm Retrieved 03 August, 2015.

Park, S., Freeman, J., Middleton, C., Allen, M., Eckermann, R., and Everson, R. 2015. "The Multi-Layers of Digital Exclusion in Rural Australia," *System Sciences (HICSS), 2015 48th Hawaii International Conference on*: IEEE, pp. 3631-3640.

Prieger, J. E. 2013. "The Broadband Digital Divide and the Economic Benefits of Mobile Broadband for Rural Areas," *Telecommunications Policy* (37:6), pp. 483-502.

Puschita, E., Constantinescu-Dobra, A., Colda, R., Vermesan, I., Moldovan, A., and Palade, T. 2014. "Challenges for a Broadband Service Strategy in Rural Areas: A Romanian Case Study," *Telecommunications Policy* (38:2), pp. 147-156.

Rajabiun, R., and Middleton, C. A. 2013. "Multilevel Governance and Broadband Infrastructure Development: Evidence from Canada," *Telecommunications Policy* (37:9), pp. 702-714.

RFNSA. 2014. "Radio Frequency National Site Archive." http://www.rfnsa.com.au/nsa/index.cgi Retrieved 10th June 2014.

Rice, R. E., and Katz, J. E. 2003. "Comparing Internet and Mobile Phone Usage: Digital Divides of Usage, Adoption, and Dropouts," *Telecommunications Policy* (27:8), pp. 597-623.

Srinuan, P., Srinuan, C., and Bohlin, E. 2012. "Fixed and Mobile Broadband Substitution in Sweden," *Telecommunications Policy* (36:3), pp. 237-251.

Straub, D., Boudreau, M., and Gefen, D. 2004. "Validation Guidelines for Is Positivist Research," *Communications of the Association for Information Systems* (13:24), pp. 380-427.

Talukder, Z., Islam, S., Mahjabeen, D., Ahmed, A., Rafique, S., and Rashid, M. 2013. "Cell Coverage Evaluation for Lte and Wimax in Wireless Communication System," *World Applied Sciences Journal* (22:10), pp. 1486-1491.

Titch, S. 2013. "The Effect of Satellite and Wireless on Rural Universal Service Policy." Los Angeles: Reason Foundation, p. 11.

Venkatesh, V., Brown, S., and Bala, H. 2013. " Bridging the Qualitative-Quantitative Divide: Guidelines for Conducting Mixed Methods Research in Information Systems," *MIS Quarterly* (37:1), pp. 21-54.

Willis, S., and Tranter, B. 2006. "Beyond the 'Digital Divide' internet Diffusion and Inequality in Australia," *Journal of sociology* (42:1), pp. 43-59.

World Bank. 2012. "Broadband Strategies Handbook: Chapter 1 Building Broadband," T.K.a.C.M. Rossotto (ed.). World Bank, p. 401.